\documentclass[aps,twocolumn,showpacs,preprintnumbers,floatfix, pre]{revtex4-2}

\usepackage{graphicx}
\usepackage{dcolumn}
\usepackage{bm}
\DeclareMathAlphabet{\mathpzc}{OT1}{pzc}{m}{it}

\usepackage[caption=false]{subfig}
\usepackage{amssymb}
\usepackage{amsmath}
\usepackage{graphicx,bm}
\usepackage{verbatim}
\usepackage{xcolor, soul}
\sethlcolor{yellow}
\usepackage{relsize}

\usepackage{float}

\begin{document}

\title{Impact of network assortativity on disease lifetime in the SIS model of epidemics
}
\author{Elad Korngut} \author{Michael Assaf}

\affiliation{Racah Institute of Physics, Hebrew University of Jerusalem, Jerusalem 91904, Israel }

\begin{abstract}
To accurately represent disease spread, epidemiological models must account for the complex network topology and contact heterogeneity. Traditionally, most studies have used \textit{random} heterogeneous networks, which ignore correlations between the nodes' degrees. Yet, many real-world networks exhibit \textit{degree assortativity}—the tendency for nodes with similar degrees to connect. Here we explore the effect degree assortativity (or disassortativity) has on long-term dynamics and disease extinction in the realm of the susceptible-infected-susceptible model on heterogeneous networks. We derive analytical results for the mean time to extinction (MTE) in  assortative networks with weak heterogeneity, and show that increased assortativity reduces the MTE and that assortativity and degree heterogeneity are interchangeable with regard to their impact on the MTE. Our analytical results are verified using the weighted ensemble numerical method, on both synthetic and real-world networks. Notably, this method allows us to go beyond the capabilities of traditional numerical tools, enabling us to study rare events in large assortative networks, which were previously inaccessible.
\end{abstract}


\maketitle

\section{\label{sec:Intro}INTRODUCTION}
Networks provide a theoretical framework for describing dynamical processes that occur in structured populations. In this realm, the underlying dynamics are largely governed by the connectivity pattern of the network---the set of links which connect the network's nodes---which is strongly influenced by whether the network is \textit{assortatively}-mixed or \textit{disassortatively}-mixed. In the former case high-degree nodes tend to connect to other high-degree nodes and vice versa, while in the latter case high-degree nodes tend to connect to low-degree nodes~\cite{newman_2002_asso_prl}. Examples for assortative networks are social networks, metabolic pathways, and scientific collaborations, while technological networks tend to be disassortative~\cite{newman_2002_asso_prl,newman_2003_asso_pre,Pastor_Satorras_2001_Dynamical_and_Correlation_Properties_of_the_Internet}. In contrast,   random networks, in which connections between nodes are  independent on their degree, and which are simpler to tackle, remain widely used as null models for studying dynamical processes~\cite{Castellano_2015_anneald_net}. This is despite the fact that correlations between neighboring nodes are ubiquitous in real-world networks,  and thus, accounting for their impact is key for understanding the  dynamics of infectious diseases within a population network~\cite{Goltsev_2008_Percolation_on_correlated_networks,VanMieghem2010_Influence_of_assortativity_and_degree_preserving_rewiring_on_the_spectra_of_networks,Bogu_2003_Absence_of_Epidemic_Threshold_in_Scale_Free_Networks_with_Degree_Correlations}.

Beside the network topology, understanding and controlling the spread of infectious diseases requires epidemiological models, which allow   mimicking disease transmission dynamics, predicting potential outbreaks, and assessing  intervention strategies~\cite{Hethcote_2000_background_inf_models,Vespignani_2001_scale_free_networks,vespignani_2001_sis_formulation,Vespignani_Pastor_2002_epidemic_scale_free,keeling2005networks,Dorogovtsev_2008_percolation,keeling2011modeling,Castellano_2015_anneald_net}. The crux of these models is to provide a rigorous framework to analyze complex interactions between various compartments representing different disease states, such as susceptible,
infected, or recovered individuals. One of the simplest compartmental models is the susceptible-infected-susceptible (SIS) model, which partitions individuals into two compartments: susceptibles (S) and infected (I)~\cite{Vespignani_2001_scale_free_networks,vespignani_2001_sis_formulation,keeling2005networks,Dorogovtsev_2008_percolation,keeling2011modeling,moore2000epidemics,karrer2010message,Pastor-Satorras_2018_sis_backtracking}. In this model susceptibles can get infected upon encountering an infected individual, while the latter can recover and become susceptible again.
Notably, modelling disease transmission
on heterogeneous networks provides a clear departure
from classic compartmental models, as in general, nodes with different degrees have different susceptibility and infectiousness. Here, quantities such as disease prevalence and epidemic threshold can be analyzed using the so-called heterogeneous mean-field approach~\cite{Castellano_2015_anneald_net}; yet, analyzing rare events such as disease extinction is much more involved, and requires going beyond mean field and accounting for demographic noise. 

Indeed, even for homogeneous population networks, where each node has the same degree, the long-time dynamics of the SIS model is nontrivial. Here, provided that the infection rate is above some threshold (see below), in the limit of an infinite population the system will converge into a stable \textit{endemic} state, in which the disease persists forever. Yet, for finite populations, this state becomes metastable, and demographic noise will eventually drive the system to extinction; that is, when the population is finite the disease lifetime is always finite, and disease extinction occurs with unit  probability~\cite{ovaskainen_2001,Assaf_2010_pre_mickey,Assaf_2017_wkb_mickey}. Notably, in a homogeneous setting the problem is effectively one dimensional, as each node has the same degree. In this case, the quasi-stationary distribution (QSD) around the (metastable) endemic state and the mean time to extinction (MTE)---the mean time it takes the system to transition from the  endemic to the extinct state--- can be analytically computed~\cite{dykman_1994_wkb_exp,ovaskainen_2001,Assaf_2010_pre_mickey,assaf2010large}. Yet, predicting rare events in heterogeneous population networks, in which nodes have varying degrees, is notoriously difficult due to the high dimensionality and complex coupling between degrees of freedom. Here,  thus far, analytical results have only been obtained  for random networks in simplified scenarios: close to the bifurcation limit, and for weakly-heterogeneous networks with undirected or directed links~\cite{Hindes_2016_jason-ira-paths,Clancy_2018_Persistence,Clancy_2018_Precise_estimates,Hindes_assaf_2019_degree_dispersion,korngut_2022_directed_sis}. Yet, the effect of assortative mixing on rare events remains unexplored.

In this work we  study, both analytically and numerically, long-time dynamics and disease extinction in the realm of the SIS model on heterogeneous, (dis)assortative  networks.  The assortativity feature is captured by the conditional probability $P(k'|k)$ that a node of degree $k$ is connected with a node of degree $k'$. In random networks this probability is independent on $k'$, whereas in (dis)assortative networks it depends on both $k$ and $k'$. Our theoretical framework is based on writing down the set of master equations for the various node types of the network and treating this equation set semi-classically.  Numerically computing the MTE on large heterogeneous networks requires highly efficient numerical algorithms. Recently, we have shown that for random heterogeneous networks, computing the MTE using the  weighted ensemble (WE)~\cite{korngut_2022_directed_sis,korngut_2024_we} method is vastly superior to standard methods such as the kinetic Monte Carlo (KMC) method~\cite{GILLESPIE1976_second_paper,Gillespie_1977,Huber_1996_we_paper,gillespie_2007_review,Keeling_2008_sim_master,Donovan_2013_we_bio_chem_reaction}. To account for degree-degree correlations, we  extend the WE method to enable dealing with (dis)assortative networks, and compute the MTE and QSD in previously inaccessible parameter regions.

The remainder of the paper is organized as follows: In Sec.~\ref{sec:theory} we introduce a model for (dis)assortative  networks and develop an analytical expression for the MTE for weakly-heterogeneous networks. In Sec.~\ref{sec:numeric} we present the numerical scheme used to calculate the MTE and to construct  assortative networks. Section~\ref{sec:results} is dedicated to presenting the results, whereas  Sec.~\ref{sec:discussion} provides a summary and discusses avenues for further research.

\section{\label{sec:theory}THEORETICAL FORMULATION}
\subsection{Model}
We begin by formulating the SIS model within a heterogeneous network topology featuring assortativity. In this model, nodes represent individuals in an isolated population of size $N$, each can be in one of two states: susceptible  ($S$) or infected  ($I$)  \cite{bailey_1993_sis_book,Pastor-Satorras_2002_Epidemic_spreading_correlated_networks,Morita_2021_sis_degree-correlated_networks}. The network's structure is captured by an adjacency matrix  $\textbf{A}$,  where an entry  $A_{ij}=1$ indicates a link between nodes $i$ and $j$ and $A_{ij}=0$ otherwise. Susceptible individuals can become infected through these links at a rate \( \lambda \) per link, while infected individuals spontaneously recover to the susceptible state at a rate \( \gamma \) per node. 

We define the degree distribution \( P(k) \), which specifies the fraction of nodes with degree \( k \). The number of nodes with degree \( k \) is denoted by \( N_k \), such that \( \sum_k N_k = N \) and \( N_k = N P(k) \). Degree correlations are encoded in the conditional probability \( P(k'|k) \), representing the likelihood of a node with degree \( k \) being connected to a node with degree \( k' \). Notably, in networks with assortative mixing various expressions of $P (k'|k)$ are possible, and we will henceforth focus on a specific choice, see below.
 
Analytical progress is possible by assuming the
annealed network approximation, i.e., a mean-field approximation over an ensemble of networks~\cite{Castellano_2015_anneald_net}. Under this approximation one replaces the adjacency matrix $\mathbf{A}$  with its expectation value $\langle\mathbf{A}\rangle$ over an 
ensemble of networks. For  uncorrelated networks, one has $P(k'|k) = k'P(k')/\left<k\right>$, and thus,
$\left<A_{ij}\right>= k_i k_j/(N\langle k\rangle)$, where $k_i$ and $k_j$ are the degrees of nodes $i$ and $j$ and $\langle k\rangle$ is the average degree.
In contrast, for assortative networks, one has to account for the nontrivial two-node  conditional probability $P(k_i|k_j)$. As a result, the ensemble-averaged adjacency matrix reads~\cite{Mendes_2002_Evolution_of_networks,Dorogovtsev_2008_percolation,Castellano_2009_Langevin_approach,Castellano_2015_anneald_net}
\begin{equation}
    \left<A_{ij}\right>=\frac{k_j P\left( k_i|k_j \right)}{N P\left(k_i\right)}.
    \label{eq:anneald_approx}
\end{equation}
For concreteness we focus henceforth on a specific choice for the conditional probability, given by~\cite{Castellano_2015_anneald_net}
\begin{equation}
\label{eq:conditional_prob}
    P(k'|k) = \frac{k'P(k')}{\left<k\right>}(1-\alpha)+\alpha \delta_{k,k'},
\end{equation}
where $\alpha$ represents the assortativity between nodes of the same degree \cite{leibenzon2024heterogeneity}. 
We now derive an expression for the mean time to extinction (MTE) on degree-heterogeneous, assortative networks. For simplicity we start with bimodal networks and then discuss how the results can be generalized to arbitrary degree distributions. 

Bimodal networks have two distinct degree classes, $k_1$ and $k_2$, with $I_i$ ($i=1,2$) denoting the number of infected nodes of degree $i$. We further assume a symmetric degree distribution, $P(k) = (\delta_{k_1, k} + \delta_{k_2, k}) / 2$. Here, each degree can be  expressed as $k_1 = k_0 (1 - \epsilon)$ and $k_2 = k_0 (1 + \epsilon)$, where $k_0=\langle k\rangle$ is the mean of the degree distribution, while $\epsilon= \sigma/k_0$ is its coefficient of variation (COV), with $\sigma$ being the standard deviation.   Thus, the infection and recovery processes for the two types of nodes  read
 \begin{eqnarray}
\label{eq:transition_rate_k_case}
    I_{i}\xrightarrow{W_{i,+}} I_{i}+1,\quad   I_{i}\xrightarrow{W_{i,-}} I_{i}-1;\quad \text{i=1,2}. 
\end{eqnarray}
Here, the recovery rate reads $W_{i,-}=I_i$, and time is measured in units of $\gamma^{-1}$. Yet, the infection rate is more involved since as \(\alpha\) increases, nodes with degree $k_i$ are increasingly likely to connect with other nodes of the same degree. Thus, the infection rate of a degree-$k_i$ node satisfies: $W_{i,+}=\lambda k_0(1\!-\!\epsilon) \left( N/2\!-\!I_{i} \right)\left[(1\!-\!\alpha)\Phi\left(x_{1},x_{2}\right)\!+\!\alpha I_{i}\right]$, where $\Phi(x_1,x_2)=\left(k_1 x_1+k_2 x_2\right)/(k_1+k_2)$ is the probability to be connected to an infected node in an uncorrelated network, and $x_{i}=I_{i}/(N/2)$  is the density of infected degree-$i$ nodes. This infection rate combines random and fully-assortative networks with relative weights $1-\alpha$ and $\alpha$ respectively, where $0\leq \alpha\leq 1$. That is, for $\alpha=0$ we recover the infection rate in  a random network, while for $\alpha=1$ we have two fully connected degree-$k_1$ and $k_2$ sub-networks that are disconnected from each other.  With these rates at hand, in the absence of demographic fluctuations, the deterministic rate equations read~\cite{Pastor-Satorras_2002_Epidemic_spreading_correlated_networks}
\begin{eqnarray}
\label{eq:rate_eq_k_case_bimodal}
    &\dot{x_{1}} = \lambda k_1\left[ \Phi(x_1,x_2)(1-\alpha)+\alpha x_1\right](1-x_1)-x_1 ,\nonumber\\
    & \dot{x_{2}} = \lambda k_2\left[ \Phi(x_1,x_2)(1-\alpha)+\alpha x_2\right](1-x_2)-x_2.
\end{eqnarray}
 At steady state, Eq.~\eqref{eq:rate_eq_k_case_bimodal} features an unstable extinct state $x_1=x_2=0$, and a stable endemic state denoted by \( \mathbf{x}^{*}=(x_1^*,x_2^*) \). A transcritical bifurcation occurs at the epidemic threshold, when $\lambda=\lambda_c$, see below. As a result it is convenient to define a basic reproduction number \( R_0 \equiv \lambda / \lambda_c \), for which a bifurcation occurs at $R_0=1$.  In a homogeneous setting, when each node has exactly $k_0$ neighbors, it is easy to show that $R_0=\lambda k_0$ (assuming the recovery rate $\gamma=1$) such that $\lambda_c=1/k_0$.

However, for heterogeneous scenarios ($\epsilon>0$), the network's degree distribution and degree assortativity affect the epidemic threshold. To account for those, several approaches for calculating $\lambda_c$ were introduced. The simplest one is the heterogeneous mean-field theory.  Here, the connectivity matrix is defined as  \( C_{kk'} = kP(k'|k) \) and the epidemic threshold becomes $\lambda_c=1/\Upsilon^{(1)}$, with $\Upsilon^{(1)}$ denoting the largest eigenvalue of the connectivity matrix~\cite{Pastor-Satorras_2002_Epidemic_spreading_correlated_networks}. The resulting reproductive number reads,
\begin{equation}
    R_0 = \lambda\Upsilon^{(1)},
    \label{eq:reproductive_number_hmf}
\end{equation}
which can be calculated numerically for a given $C_{kk'}$.
Analytical progress can be made in specific cases~\footnote{For generic assortative networks with arbitrary heterogeneity, 
the epidemic threshold $\lambda_c$ is given by the inverse of the largest eigenvalue $\Lambda^{(1)}$ of the adjacency matrix $\textbf{A}$, such that 
$R_0 = \lambda\Lambda^{(1)}$. Yet, this additionally requires a large spectral gap between the first and second eigenvalue, such that $\Lambda^{(1)}\gg\Lambda^{(2)}$~\cite{Pastor-Satorras_2019_assortative_power-law_networks,Mata_2013_quenched_mean_theory,Chakrabarti_2008_Epidemic_thresholds_in_real_networks}, which is often satisfied for strongly-connected networks~\cite{,hindes_2017_paths_in_complex_networks_vector_centrality}.  Notably, for the networks examined in this work, working with either  the adjacency or connectivity matrix yields identical epidemic thresholds and endemic states, given \( R_0  \).
}. In the simple case of zero assortativity, the connectivity matrix becomes $C_{kk'}=kk'P(k')/\left<k\right>$, and the largest eigenvalue reads: $\Upsilon^{(1)}=\left<k^2\right>/\left<k\right>$, which yields the known result in random heterogeneous networks: $R_0\! =\! \lambda \left<k^2\right>\!/\!\left<k\right>$ \cite{Boguna_2003_absence_correlation}.

Another analytically tractable case is an assortative bimodal network. Here, the connectivity matrix reads
\[
\mathbf{C}=\frac{1}{2k_0}\begin{bmatrix}

k_{1}^{2}(1-\alpha)+\alpha k_{1} & k_{1}k_{2}(1-\alpha) \\
k_{1}k_{2}(1-\alpha) & k_{2}^{2}(1-\alpha)+\alpha k_{2}
\end{bmatrix}.
\]
Computing $\Upsilon^{(1)}$ and plugging it into Eq.~\eqref{eq:reproductive_number_hmf}  yields
\begin{eqnarray}
&&R_0=\frac{\lambda k_0}{2} \Bigl(1 + \alpha + \epsilon^2 - \alpha \epsilon^2 +\nonumber \\
&&\left. \sqrt{( \alpha-1)^2 - 2 \left[( \alpha -2) \alpha -1 \right] \epsilon^2 + (1 - \alpha)^2 \epsilon^4}\right).
\label{eq:reproductive_bimodal}
\end{eqnarray}
In the random-network limit, $\alpha\!\to\! 0$, one has $R_0\!=\!\lambda k_0(1\!+\!\epsilon^2)$, which equals $\lambda \langle k^2\rangle/\langle k\rangle$. Another interesting limit is when the heterogeneity is weak, $\epsilon\ll 1$ in which case one finds: $R_0\simeq \lambda k_0[1+\epsilon^2/(1-\alpha)]$. Notably, as long as $R_0>1$ the endemic state persists, and in the limit of an infinite population size, starting from any nonzero infection size the epidemic will remain forever in the endemic state. 

Yet, for finite populations demographic noise emanating from the discreteness of individuals and stochasticity of reactions may be significant. Here, even if \( R_0 > 1 \), disease extinction occurs with finite probability. To account for demographic noise we write down a master equation for the joint probability to find $I_1$ and $I_2$ number of infected individuals on degree-$k_1$ and $k_2$ nodes, at time $t$:
\begin{eqnarray}
        &\dot{P}_{I_1,I_2}(t)=\left\{\lambda k_1(E_{I_1}^{-1}\!-\!1)(N/2\!-\!I_1)\left[\Phi(I_1,I_2)(1\!-\!\alpha)+\right. \right. \nonumber\\
        &\left.\left.\alpha I_1\right] +\lambda k_2(E_{I_2}^{-1}-1)(N/2-I_2)\left[\Phi(I_1,I_2)(1-\alpha)\right.\right. \nonumber\\
        &\left. +\left.\alpha I_2\right] +(E_{I_1}^{1}-1)I_1+(E_{I_2}^{1}-1)I_2 \right\}P_{I_1,I_2},
    \label{eq:master_bimodal}
\end{eqnarray}
with a step operator $E_{I}^{j}\!=\!f(I\!+\!j)$  used for brevity~\footnote{In the general case of an arbitrary heterogeneous network with degrees  $1,\dots,k_{\text{max}}$, the master equation can be written in a similar manner to Eq.~(\ref{eq:master_bimodal}) for the probability \( P(\mathbf{I}, t) \) of finding \( \mathbf{I} = \{I_1, \ldots, I_{k_{\text{max}}}\} \) infected individuals on nodes of degrees $1,\ldots,k_{\text{max}}$. Here, the infection rate satisfies
$W_{k}^{+}(\mathbf{I})=(\lambda k/N) (N_k-I_k) \sum_{k'=1}^{k_{\text{max}}} P\left(k'|k\right)I_{k'}$, with $I_{k}$  being the number of infected on degree-$k$ nodes, while the recovery rate satisfies $W_{k}^{-}(\mathbf{I})=\gamma I_k$.}.

While Eq.~(\ref{eq:master_bimodal}) does not admit an exact solution, one can use the  Wentzel-Kramers-Brillouin (WKB) approximation in the limit of $N\gg 1$ to transform the set of master equations into a Hamilton-Jacobi equation. In this limit, after a short time transient the system enters  a long-lived metastable endemic state and stays there for a long time before eventually going extinct. During this stage, the probability of remaining in the endemic state decreases slowly over time, while the likelihood of extinction increases. Thus we use the metastable ansatz $P_{I_1,I_2}=\pi_{I_1,I_2}e^{-t/\tau}$ where $\tau$ is the exponentially large MTE, see below, and $\pi_{I_1,I_2}$ is the QSD~\cite{dykman_1994_wkb_exp,Assaf_2010_pre_mickey,Assaf_2017_wkb_mickey,Hindes_assaf_2019_degree_dispersion}. Further employing the WKB approximation on the QSD, $\pi_{I_1,I_2} \equiv\pi\left(x_1,x_2\right) \sim\exp \left[-N \mathcal{S}\left( x_1,x_2\right) \right] $, with \( \mathcal{S}(x_1,x_2) \) denoting the action function, and substituting it into Eq.~\eqref{eq:master_bimodal}, results in a stationary Hamilton-Jacobi equation, $H\left(x_1,x_2,\partial_{x_1} \mathcal{S},\partial_{x_2} \mathcal{S}\right)=0$,  with a Hamiltonian
\begin{eqnarray}
\label{eq:hamiltonian_bimodal}H\left(x_1,x_2,p_1,p_2 \right) &&=\frac{x_1}{2}\left( e^{-p_1}-1 \right)+\frac{x_2}{2}\left(e^{-p_2}-1\right)\\
    &&\hspace{-15mm}+\frac{\lambda k_1}{2}[(1-\alpha)\Phi(x_1,x_2)+\alpha x_1](1-x_1)\left(e^{p_1}-1\right) \nonumber\\
    &&\hspace{-15mm}+\frac{\lambda k_2}{2}[(1-\alpha)\Phi(x_1,x_2)+\alpha x_2](1-x_2)\left(e^{p_2}-1\right),\nonumber
\end{eqnarray}
where  $p_i/2=\partial_{x_i} \mathcal{S}$  is the normalized momenta of group $i$. Solving the associated Hamilton's equations, $\dot{x}_i\!=\!\partial_{p_i}H/2$ and $\dot{p}_i \!=\!-\partial_{x_i}H/2$, allows finding the action function $S(x_1,x_2)\!=\!\int (p_1/2)dx_1 \!+\! (p_2/2)dx_2$~\cite{dykman_1994_wkb_exp,Assaf_2010_pre_mickey,Assaf_2017_wkb_mickey,Hindes_assaf_2019_degree_dispersion}. Notably, a similar procedure can be done in principle for generic networks with arbitrary heterogeneity and assortativity~\footnote{For generic heterogeneous networks, one obtains: $H(\mathbf{x},\mathbf{p})\!=\!\sum_{k=1}^{k_{\text{max}}}\!P(k)\left[ w_k^{+}(e^{p_k} \!-\!1)\!+\!x_k( e^{-p_k}\!-\!1)\right]$. Here, $p_k=\partial_{x_k} \mathcal{S}/P(k)$ is the normalized momenta of group $k$ with $\mathcal{S}$ being  the action, $w_k^+\!=\!\lambda k (1\!-\!x_k)[\alpha x_k \!+\!( 1\!-\!\alpha) \Phi(\mathbf{x})]$ is the infection rate, and $\Phi(\mathbf{x})\!=\!\sum_{k{'}}\!k' P(k')x_{k'}\!/\!\left<k\right>$~\cite{Hindes_2016_jason-ira-paths}.}.

\subsection{Action barrier for weak heterogeneity}
While these Hamilton's equations can always be solved numerically, for any choince of $\epsilon$, analytical progress can be made in the limit of weakly-heterogeneous networks, where $\epsilon\ll 1$. In this limit it is convenient to apply the following canonical transformation and introduce the new variables: \( u = (x_{1} - x_{2}) / 2 \),  \( p_u = p_{1} - p_{2} \), \( w = (x_{1} + x_{2}) / 2 \), and \( p_w = p_{1} + p_{2} \). As we will show, in the small $\epsilon$ limit one obtains a time scale separation between $w$ and $p_w$ and $u$ and $p_u$. 
First we look for the fixed points, which are determined by solving the Hamilton's equations at equilibrium. Doing so, one can show that system has an endemic fixed point at \((w^*,u^*,0,0)\), and an extinct fixed point at  \((0,0,p_{w}^{*},p_{u}^{*})\), both of which are saddles: 
\begin{eqnarray}
u^{*} &&= -\frac{x_0}{\left(R_0-\alpha  \right)}\epsilon,\quad p_{u}^{*} = \frac{2 (R_0-1) }{R_0-\alpha }\epsilon , \nonumber \\
w^{*}&&=x_0 \left[1-\frac{ (2-\alpha ) R_0-\alpha }{(1-\alpha ) (R_0-\alpha )^2}\epsilon ^2\right] , \nonumber \\
p_{w}^{*}&&=-2 \ln (R_0) + \frac{(R_0-1)\left[1+(3-\alpha) R_0-3 \alpha\right]}{(1-\alpha) (R_0-\alpha )^2} \epsilon ^2,
\label{eq:fixed_point_low_disperssion}
\end{eqnarray} 
where $x_0=1-1/R_0$  is the endemic solution in the well-mixed case. Since the above transformation is canonical, the action barrier and resulting MTE satisfy
\begin{equation}
\quad \tau\sim e^{N \Delta\mathcal{ S}},\quad  \Delta\mathcal{S}= \frac{1}{2}\int p_{w}dw+\frac{1}{2}\int p_{u}du,
\label{eq:action_bimodal_canonical_cordinate_only_pw}
\end{equation}
where the integrals are calculated along the optimal path---a heteroclinic trajectory connecting the endemic and extinct fixed points~\cite{dykman_1994_wkb_exp}. 

We now find the trajectories $p_w(w)$ and $p_u(u)$ along the optimal path to extinction, in the limit of $\epsilon\ll 1$, where we  keep terms up to ${\cal O}(\epsilon^2)$, see below. We do so in the spirit of Ref.~\cite{Hindes_assaf_2019_degree_dispersion}, where a random weakly-heterogeneous network with zero assortativity was considered.

The trajectory $p_u(u)$ starts from the endemic state \(\left[u^*,0\right]\) and reaches the extinct state \(\left[0,p_{u}^{*}\right]\). Since  \( u^{*} \) and  \( p_{u}^{*} \) both scale as \( \mathcal{O}(\epsilon) \), see Eq.~(\ref{eq:fixed_point_low_disperssion}), we  thus  expect both $u$ and $p_u$ to scale as
${\cal O}(\epsilon)$ along the entire path. As a result, since the integral over $p_udu$
already scales as ${\cal O}(\epsilon^2)$, it is suffcient to approximate
$p_u(u)$ as a straight line connecting \(\left[u^*,0\right]\) and  \(\left[0,p_{u}^{*}\right]\), which yields:
\begin{equation}
     p_u(u) = 2 R_0 u + \frac{2 (R_0-1) }{R_0-\alpha }\epsilon.
     \label{eq:pu_bimodal_low_disperssion}
 \end{equation}

In order to find the trajectory $p_w(w)$,  we express Eq.~\eqref{eq:hamiltonian_bimodal} using the canonical transformation $x_1 = w + u$, $x_2 = w - u$,
$p_1 = (p_w + p_u)/2$ and $p_2 = (pw - pu)/2$. Under the assumption that $u$ and $p_u$ scale as $\mathcal{O}(\epsilon)$, the resulting Hamiltonian reads in the leading order in $\epsilon$:
\begin{equation}
    H(w,p_w,u,p_u)\!=\!2w(e^{p_w/2}\!-\!1)\!\!\left[R_0(1\!-\!w)\!-\!e^{-p_w/2}\right]+{\cal O}(\epsilon^2).
    \label{eq:hamilton_low_eps_leading}
\end{equation}
As a result, we find in the leading order $p_{w}^{(0)}(w) = -2 \ln[w(1 - R_0)]$. 
To find the  ${\cal O}(\epsilon^2)$ subleading correction, we notice that the corrections to $w^*$ and $p_w^*$ in Eqs.~\eqref{eq:fixed_point_low_disperssion} scale as $\mathcal{O}(\epsilon^2)$. We thus look for the perturbed solution in the form:  $p_w(w)=p_{w}^{(0)}\left[w\left(1+\phi \epsilon^2\right)\right] +\left[p_{w}^{*}-p_{w}^{(0)}\right]\left(1-w/w^*\right)$, where $\phi$ is a parameter yet to be determined, and $p_{w}^{*}$ and $w^{*}$ are given by Eqs.~(\ref{eq:fixed_point_low_disperssion}) with $p_{w}^{*}-p_{w}^{(0)}(0)\sim \epsilon^2$.  It is straightforward to verify that at $w=0$, $p_w(0)=p_{w}^{*}$. At $w=w^*$, we demand that $p_w(w^*)=0$ up to $\mathcal{O} (\epsilon^4)$ corrections. This yields $\phi = \left[\alpha +(\alpha -2) R_0\right]/\left[(\alpha -1) (R_0-\alpha )^2\right]$, which reduces to $\phi=2/R_0$ in the random network case with $\alpha=0$~\cite{Hindes_assaf_2019_degree_dispersion}. Rewriting the trajectory $p_w(w)$ in powers of $\epsilon$ we find
\begin{eqnarray}
    p_w(w) &&= p_{0}^{(w)}(w)-\frac{1}{(\alpha -1) (R_0-\alpha )^2}\left\{ \frac{2 w \left[\alpha \!+\!(\alpha \!-\!2) R_0\right]}{w-1} \right.\nonumber\\
    &&\Biggl.+\left[1 + R_0 ( w - 1)\right] \left[ 3 \alpha + R_0 (\alpha -3) -1 \right] \biggl\} \epsilon^2.
    \label{eq:pw_path_low_disperssion}
\end{eqnarray}
Putting everything together, and performing the integration in Eq.~\eqref{eq:action_bimodal_canonical_cordinate_only_pw} using the trajectories found in Eq.~\eqref{eq:pu_bimodal_low_disperssion} and Eq.~\eqref{eq:pw_path_low_disperssion} leads to the following action barrier
\begin{eqnarray}
    \Delta S &=& S_0 - \left\{\frac{(R_0 - 1) \left[5 \alpha - (\alpha - 3) (R_0 - 4) R_0 + 1\right]}{4 R_0 (1 - \alpha) (R_0 - \alpha)^2} \right. \nonumber \\
    && \Biggl. + \frac{\left[(2 - \alpha) R_0 - \alpha \right] \ln(R_0)}{(1 - \alpha) (R_0 - \alpha)^2} \biggl\} \epsilon^2 + \mathcal{O}(\epsilon^4),
    \label{eq:action_assortative_low_dispersion}
\end{eqnarray}
where $S_0=1/R_0+\ln R_0-1$ is the action barrier for a degree-homogeneous network. This is a central result of this work; it demonstrates the impact of assortativity on the MTE for networks with weak heterogeneity. While our analytical derivation was carried on a bimodal network, one can easily show that Eq.~(\ref{eq:action_assortative_low_dispersion}) holds for weakly-heterogeneous networks with an arbitrary degree distribution, as long as the skewness is not too large~\cite{Hindes_assaf_2019_degree_dispersion}. 

Notably, Eq.~(\ref{eq:action_assortative_low_dispersion}) shows that the action barrier monotonically decreases  with increasing assortativity and COV, implying that as assortativity and heterogeneity increase, the MTE decreases. In the limit of $\alpha\to 0$  Eq.~\eqref{eq:action_assortative_low_dispersion} reduces to
\begin{equation}
\label{eq:mj_action_correction}
   \hspace{-3mm} \Delta\mathcal{S} \simeq\mathcal{S}_{0}\! -\! \frac{(R_{0} \!-\! 1)(1\! - \!12R_{0} \!+\! 3R_{0}^2) + 8R_{0}^2\ln(R_{0})}{4R_{0}^3}\epsilon^{2}\!,
\end{equation}
which is consistent with results found in Ref. \cite{Hindes_assaf_2019_degree_dispersion}.
Another interesting limit is when $R_0-1\ll 1$, in which Eq.~(\ref{eq:action_assortative_low_dispersion}) reduces to
\begin{eqnarray}
\label{eq:smallR_action_correction}
    \Delta\mathcal{S} =\mathcal{S}_{0} - \frac{3  (R0 - 1)^2}{2(\alpha-1)^2}
\epsilon^{2}.
\end{eqnarray}
Finally, looking at the limit of very strong assortativity, $\alpha\to 1$, we see that for our perturbative theory to be valid, $\alpha$ cannot be too close to $1$. In fact, one has to demand that $1-\alpha\gg \epsilon^2$ for the correction $\Delta\mathcal{S}$ to be small compared to $S_0$. In practice, see below, we worked with networks up to $\alpha=0.7$. As stated above, as $\alpha$ approaches $1$, the network becomes disconnected.

\section{NUMERICAL METHODOLOGY \label{sec:numeric}}
To study assortative networks, in addition to the theoretical analysis we used various numerical tools that allow going beyond weakly-heterogeneous networks. The first method involved solving the set of master equations [Eq.~\eqref{eq:master_bimodal}] numerically and finding the MTE. Here, for $N\gg 1$,  the MTE turns out to be exponentially large [see Eq.~(\ref{eq:action_bimodal_canonical_cordinate_only_pw})]. In this case, it is approximately given by the inverse of the smallest (in absolute value) eigenvalue of the transition matrix from state $i$ to $j$ ($i,j\in [1,N]$)~ \cite{ovaskainen_2001,Keeling_2008_sim_master}. Although this method provides accurate solutions including pre-exponential corrections, it only outperforms "brute-force" approaches such as KMC methods (see below) when the dimensionality $d$ is low, as the matrix size scales as \(\mathcal{O}(N^{2d})\). Thus,  for bimodal networks with $d=2$ and a matrix size  of ${\cal O}(N^4)$, finding the MTE via a numerical solution of the master equation  is rather  efficient compared to KMC methods, see below. Yet, solving the master equation for  realistic networks with $d\gg 1$, is infeasible and requires other numerical schemes.

An alternative way of finding the MTE is to perform network simulations, which attempt to mimic the stochastic processes of infection and recovery. Naturally, before simulating the dynamics, the network itself has to be generated with a prescribed degree distribution $P(k)$ and degree-degree correlation function. To generate random networks with zero assortativity, one often uses the configuration model~\cite{Fosdick_2018_config_model}. Here, each node is assigned a predefined number of stubs corresponding to its degree, and these stubs are randomly paired to form links, thereby preserving the degree distribution while preventing correlations between the network's nodes. However, to create a network with a prescribed assortativity, the stubs have to be connected in a different manner.
One way is to connect a fraction $\alpha$ of stubs between nodes of the same degree, and then apply the configuration model to the remaining stubs. That is, stubs are assigned to each node based on the degree distribution $P(k)$, and each degree-\(k\) node connects $\alpha k$ of its stubs to other degree-$k$ node stubs, while the remaining $(1-\alpha)k$ stubs are then randomly paired across all nodes. This process generates an assortative network with assortativity, $\alpha>0$ \cite{Molloy_1995_random_graphs,Weber_2007_random_networks,leibenzon2024heterogeneity}. However, it does not allow the creation of disassortative networks, where $\alpha<0$.

To generate disassortative networks, one can use, e.g., the Xulvi-Brunet–Sokolov algorithm. Here, initially the network topology is generated using the configuration model. Next, network links are iteratively rewired to adjust the assortativity. Namely, two links involving four nodes are randomly selected and ordered based on their degrees. The links are then rewired such that with  probability $|\alpha|$ one link connects the highest-degree node to the lowest-degree node, whereas with complementing probability $1-|\alpha|$ the nodes are randomly connected. This procedure brings about an increase in disassortativity as $\alpha$ increases in absolute value.  If the rewiring step introduces links that already exist in the network, the step is repeated to prevent link duplication. Contrary to the previous method, this process allows for the generation of both disassortative and, with minor adjustments, assortative networks~\cite{Xulvi-Brunet_2004_Reshuffling_assortative}. However, regardless of the method used to create the network topology, both approaches produce an adjacency matrix, from which the basic reproduction number $R_0$ can be calculated. 

Once a heterogeneous network is created, we are in the position to  simulate the SIS dynamics. The conventional method  to do so is to employ the KMC method, such as the Gillespie algorithm \cite{GILLESPIE1976_second_paper,Gillespie_1977}. Here transitions between states occur at exponentially distributed waiting times, based on the infection and recovery rates. Due to the stochastic nature of the transition events, the network extinction times follow an exponential distribution. Hence, the network's MTE and its confidence bounds are found by performing multiple Gillespie simulations and fitting the extinction times to an exponential distribution. 
Finally, the overall MTE is obtained by averaging over multiple network realizations, while the standard deviation of the confidence bounds provides the MTE's error bars. Importantly, this simulation method is inherently slow, since runtime grows exponentially with population size and basic reproductive number.

The WE method on the other hand offers fast runtimes in high-dimensional systems by simulating multiple  realizations concurrently~\cite{korngut_2024_we}. First, the phase space is partitioned into bins, each containing \(m\) realizations, which are evolved over small time intervals \(\Delta t_{WE}\) using the Gillespie method, during which their states are updated. Realizations that explore previously unrecorded regions closer to the extinction state are replicated. Meanwhile, bins with an excess or shortage of realizations are resampled to maintain \(m\) realizations, with their weights adjusted accordingly. Finally, the total weight of the  extinct realizations is recorded, and the process is  repeated \(M\) times, providing estimates for  the MTE and QSD.
Notably, the stochastic nature of the WE method introduces instability into the MTE calculation. As a result, the parameters of the WE simulation, $\Delta t_{WE}$ and $m$, are varied to identify the optimal set, which minimizes variation across different WE realizations for the same network.

\begin{figure}[t]
    \includegraphics[width=0.85\linewidth]{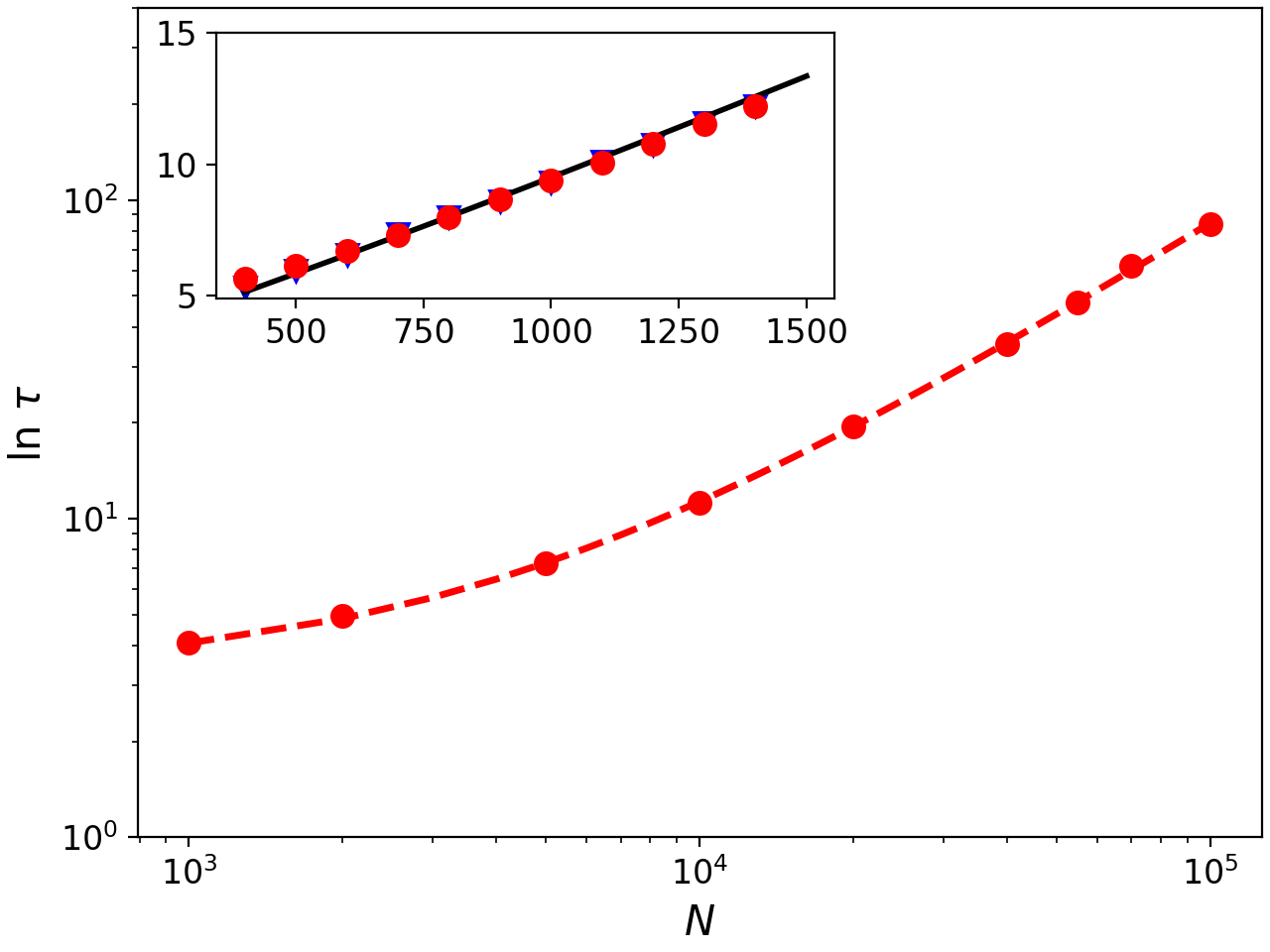}
\vspace{-3mm}    
\caption{The logarithm of the MTE, $\ln(\tau)$, versus population size \(N\), for a gamma distributed network with \(\langle k \rangle = 10\),  \(R_0 = 1.2\), \(\epsilon = 1.9\) and \(\alpha = 0.3\). Symbols represent WE simulations with parameters $m=500$, $\Delta t_{WE}=0.5$ and $M=70$, while the dashed line is the theoretical prediction (see text). The inset displays the logarithm of MTE versus $N$ for a bimodal network. Circles and triangles represent WE and KMC simulations, respectively, while the  solid line represents the solution of the master equation. Parameters here are \(\langle k \rangle = 100\), \(\epsilon = 0.7\),  \(R_0 = 1.2\), and  \(\alpha = 0.3\). }
    \label{fig1}
\end{figure}

\section{Results\label{sec:results}}
Below, we apply the WE method  to explore networks with bimodal, gamma, beta, log-normal, and inverse-Gaussian degree distributions across various values of $N$, $R_0$, $\epsilon$ and $\alpha$. As a first step, we calculated the dependence of the MTE on the network size $N$. In Fig.~\ref{fig1} we plot the natural logarithm of the MTE versus $N$ for a gamma distributed network, where the MTE is computed for network sizes up to $10^5$. In the well-mixed case, the dependence on $N$ is expected to have the following form: \( \tau \simeq A\,N^{\beta}\,e^{N{\Delta\cal S}} \), where $\Delta\mathcal{S}=S_0$ is the action barrier to extinction, $S_0$ is given below Eq.~(\ref{eq:action_assortative_low_dispersion}) and $A\,N^{\beta}$ is a subleading-order prefactor, while $A$ and $\beta$ are constants~\cite{Assaf_2010_pre_mickey,assaf2010large,Hindes_2016_jason-ira-paths}. Notably, as can be seen in the figure, this functional dependence provides an adequate theoretical prediction to the  numerical results, even in the case of assortative, heterogeneous networks. In the inset of Fig.~\ref{fig1} we show a calculation of the MTE  via three different numerical techniques, on bimodal networks. Here we compare between the numerical solution of the underlying set of master equations~(\ref{eq:master_bimodal}), KMC simulations and WE simulations. While the agreement between all methods is very good, one should note that the runtime of KMC simulations grows exponentially with $N$ (compared to a linear runtime for the WE simulations~\cite{korngut_2024_we}), and thus, the KMC method is highly inefficient for the study of rare events on large population networks.

\begin{figure}[t]
    \includegraphics[width=1.0\linewidth]{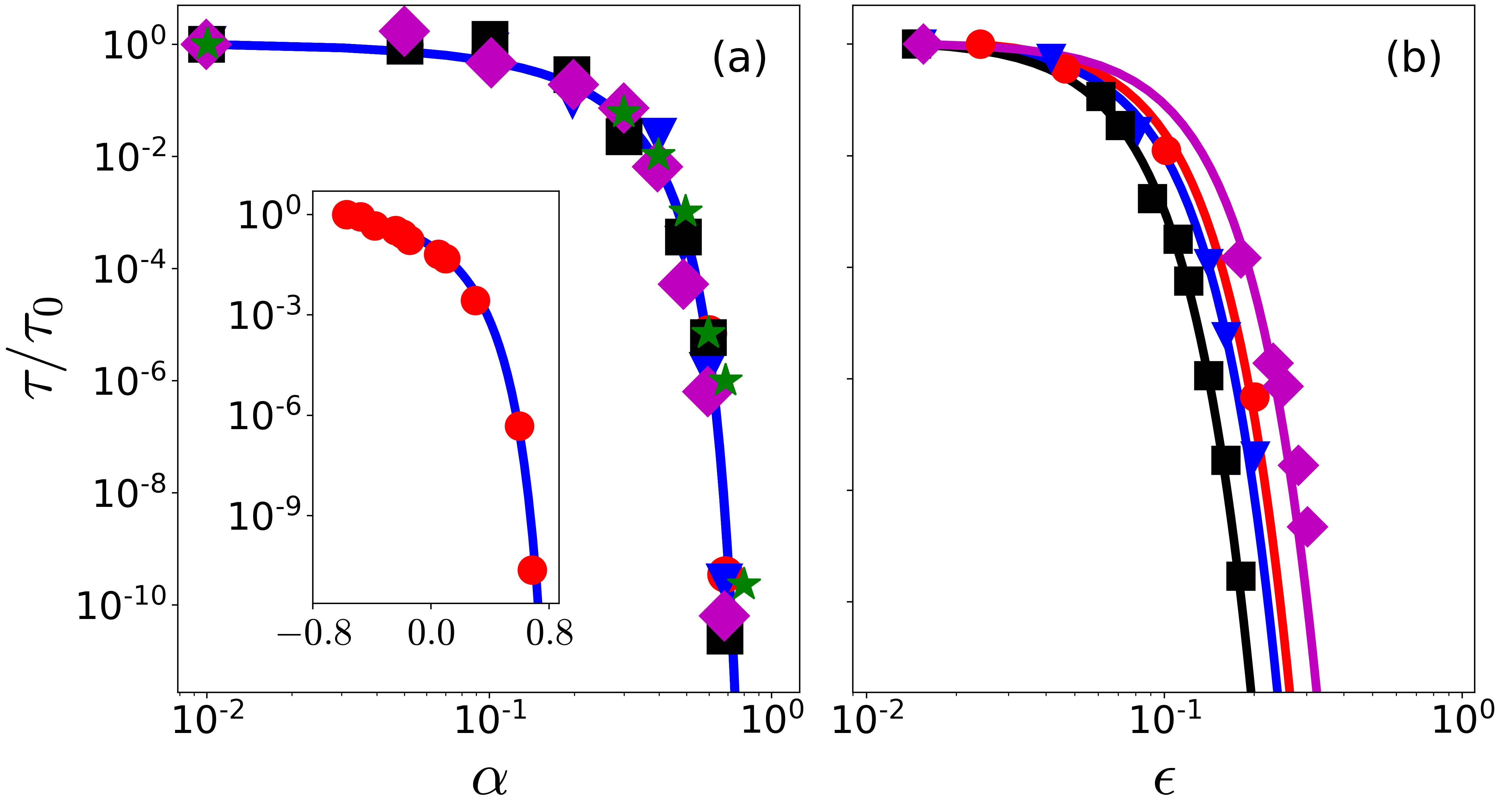}
\vspace{-5mm}    
\caption{The relative change in the MTE, $\tau/\tau_0$,  for networks with  $N = 5\times 10^3$,  \(\langle k \rangle = 50\) and \(R_0 = 1.3\), where $\tau_0$ is the MTE for a network with $\alpha=0$ (a) and $\epsilon=0$ (b). In all panels solid curves represents our analytical results~\eqref{eq:action_assortative_low_dispersion} and symbols represent WE simulations results with parameters $\tau_{WE}=0.5$, $m=500$ and $M=70$. In  (a) we plot the relative MTE on a log-log scale as function of $\alpha$ for $\epsilon=0.1$, where the circles, triangles, squares, diamonds and stars respectively represent networks with gamma, inverse-Gaussian, log-normal, beta, and bimodal distributions. The inset displays \(\tau/\tau_0\) on a semi-log scale for both assortative and disassortative gamma distributed networks. In  (b) we plot the relative MTE on a log-log scale as function of $\epsilon$ for gamma distributed networks. The different symbols represent networks with different assortativity (and disassortativity): circles for random networks with \(\alpha = 0\), triangles and squares for assortative networks with  \(\alpha = 0.1\) and \(\alpha = 0.3\) respectively, and diamonds for disassortative networks with \(\alpha = -0.3\).
 }
    \label{fig2}
\end{figure}

Once we have verified the accuracy of the WE simulations, we  studied the dependence of the MTE on the assortativity and heterogeneity strength. In Fig.~\ref{fig2} we illustrate the relative change in the MTE compared to its maximum value, \( \tau_0 \), as function of the assortativity (panel a) and heterogeneity (panel b). Here, we ran WE simulations on various assortative networks with gamma, beta, log-normal, and inverse-Gaussian degree distributions. Furthermore, the inset of panel (a) shows results for both assortative and disassortative gamma networks. Note that, since the epidemic threshold $\lambda_c$ changes as $\alpha$ or $\epsilon$ are varied, we made sure that for each separate network $\lambda$ is changed accordingly in order to maintain  a constant reproductive number $R_0$. In both panels of Fig.~\ref{fig2} one observes an excellent agreement between the WE simulations and the theoretical prediction~[Eq.~\eqref{eq:action_assortative_low_dispersion}]. Importantly, as the assortativity $\alpha$ increases, the MTE decreases, and a similar qualitative dependence appears in panel (b) for various values of $\alpha$; here, as $\alpha$ is increased, the rate of decrease of $\tau$ versus $\epsilon$ increases. The decrease of $\tau$ as the assortativity $\alpha$ increases is due to the fact that, as $\alpha$ increases the high-degree nodes are more likely to cluster. Once this high-degree cluster recovers, the effective reproductive number drastically decreases, which greatly expedites disease clearance. In contrast, as assortativity becomes negative (disassortative networks), the MTE grows as $\alpha$ (in absolute value) grows, as can be seen in the inset of panel (a). This is likely because for strongly disassortative networks, high-degree nodes are connected to many low-degree nodes, and thus, susceptible high-degree nodes tend to get re-infected much more easily. As a result, in this case disease clearance occurs with a much smaller probability as observed in the inset.

\begin{figure}[t]
    \includegraphics[width=0.84\linewidth]{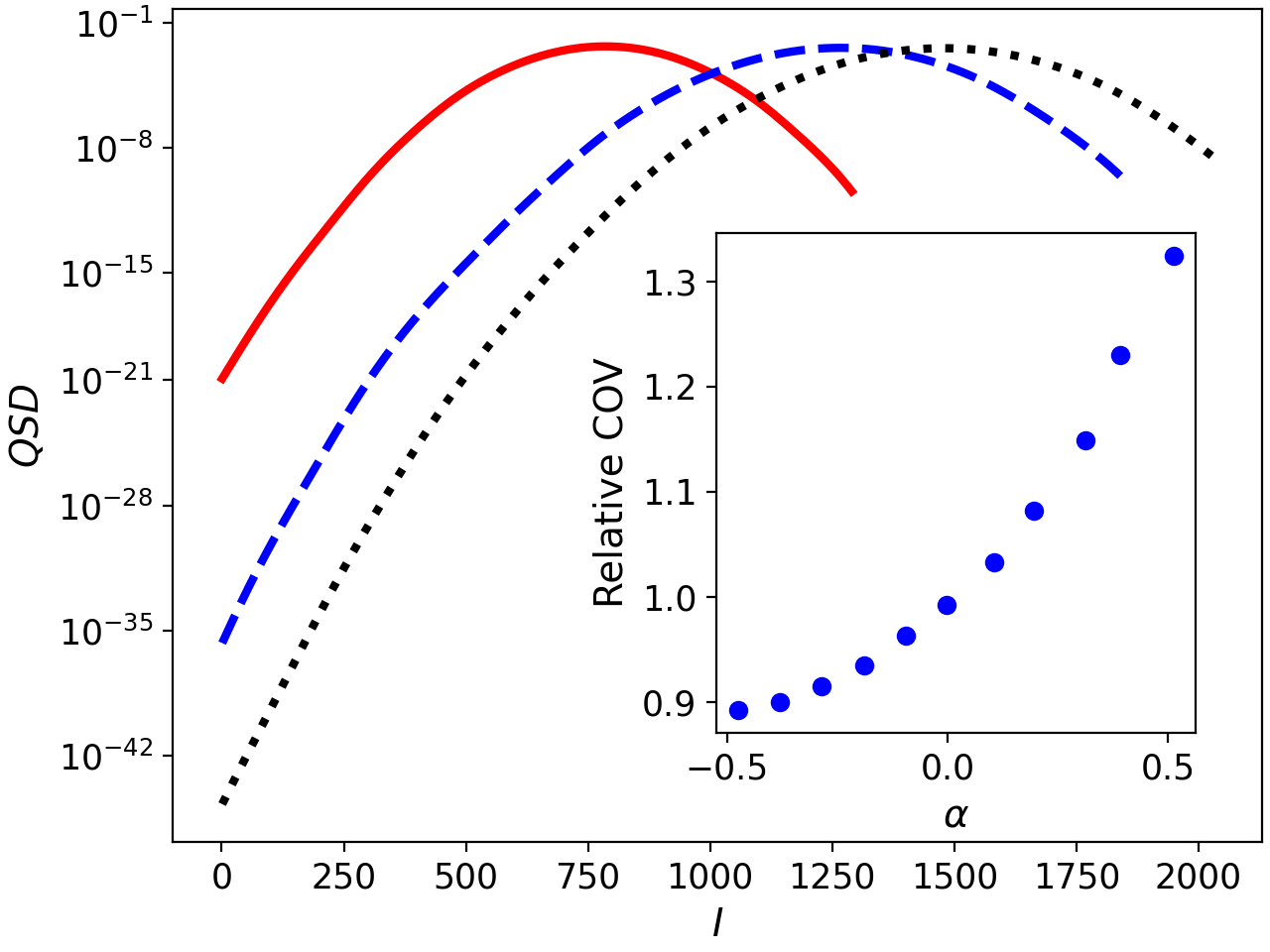}
\vspace{-3mm}    
\caption{Semi-log plot of the QSD as a function of the total number of infected individuals in the network, \( I \), for a gamma-distributed network with \( N = 10^4 \) nodes, \( R_0 = 1.24 \), \( \langle k \rangle = 50 \), and \( \epsilon = 0.5 \). Lines represent WE simulations with parameters \( m = 500 \), \( \tau_{WE} = 0.5 \), and \( M = 70 \); the dotted line corresponds to a disassortative network with \( \alpha = -0.5 \), the dashed line to a random network with \( \alpha = 0.0 \), and the solid line to an assortative network with \( \alpha = 0.5 \). Inset: relative COV of the QSD, see text,  as a function of \( \alpha \) for networks with the same parameters as in the main figure.}
    \label{fig3}
\end{figure}

These findings are further supported when comparing the QSDs for disassortative and assortative networks, while keeping the heterogeneity strength constant. In Fig.~\ref{fig3} we plot the QSD for three gamma-distributed networks, differing only in their assortative mixing. As assortativity increases (from a negative to a positive value), the disease prevalence declines, while the probability of extinction (close to \( I = 0 \)) grows. Additionally, in the inset of Fig.~\ref{fig3}, we plot the relative coefficient of variation (COV) of the QSD, which is defined as the QSD's standard deviation divided by its mean, normalized by the COV of an uncorrelated network ($\alpha=0$). One can see that as $\alpha$ increases, the fluctuations magnitude increase, which is a further indicator for more rapid extinction.

\begin{figure}[t]
    \includegraphics[width=0.80\linewidth]{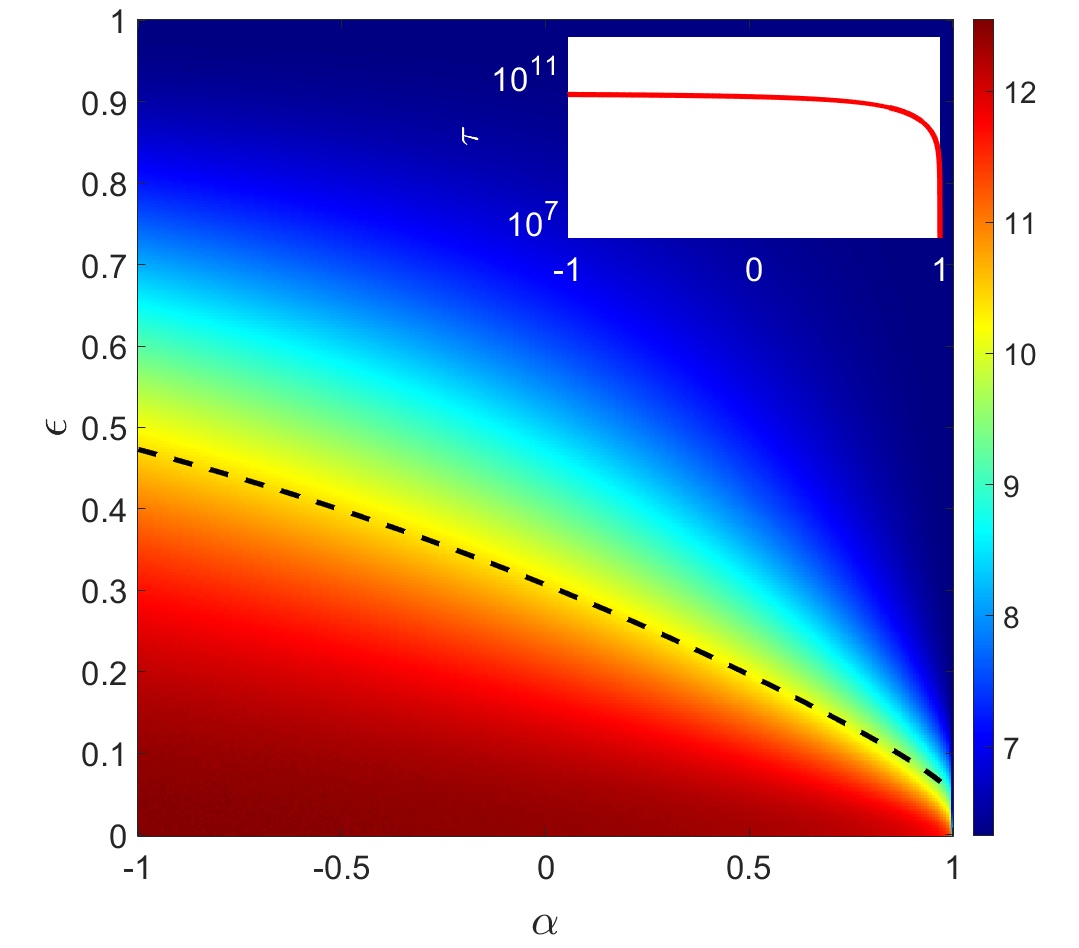}
\vspace{-4mm}    
\caption{Logarithmic heatmap of  the MTE, $\log_{10}(\tau)$, versus $\alpha$ and $\epsilon$, obtained by numerically solving  master equation~(\ref{eq:master_bimodal}) for a bimodal network with $N=400$ and $R_0=1.5$. Along the dashed line  the steady state solution to Eq.~\eqref{eq:rate_eq_k_case_bimodal}  is fixed to be $x^{*}=0.9 (1-1/R_0)$. Inset: MTE along the dashed line.} 
    \label{fig4}
\end{figure}

In Fig.~\ref{fig2} we have seen that the qualitative dependence of the MTE of $\alpha$ and $\epsilon$ is similar. To further explore this point, we plot in Fig.~\ref{fig4} a heatmap of the disease mean lifetime for a bimodal network (\(k=2\)), obtained by numerically solving master equation~(\ref{eq:master_bimodal}). Here, we see that the MTE is highest when \(\epsilon = \alpha = 0\), and as \(\epsilon\) or \(\alpha\) increase, the MTE decreases. Notably, the heatmap allows to find curves on the $\alpha-\epsilon$ plane on which the MTE is fixed. In particular, an increase in $\alpha$ can be balanced by a decrease in $\epsilon$ and vice versa. Furthermore, along the dashed line in Fig.~\ref{fig4} the  steady-state disease prevalence $x_*$---the total number of infected individuals divided by $N$---is fixed. This curve is generated by specifying a desired $x^*$ value and adjusting both \(\alpha\) and \(\epsilon\) such that the solutions to Eqs.~\eqref{eq:rate_eq_k_case_bimodal} yield the target infected density. Remarkably, the dashed line aligns closely with the regions of constant MTE, as shown in the inset of Fig.~\ref{fig4}.

\begin{figure}[t]
    \includegraphics[width=0.82\linewidth]{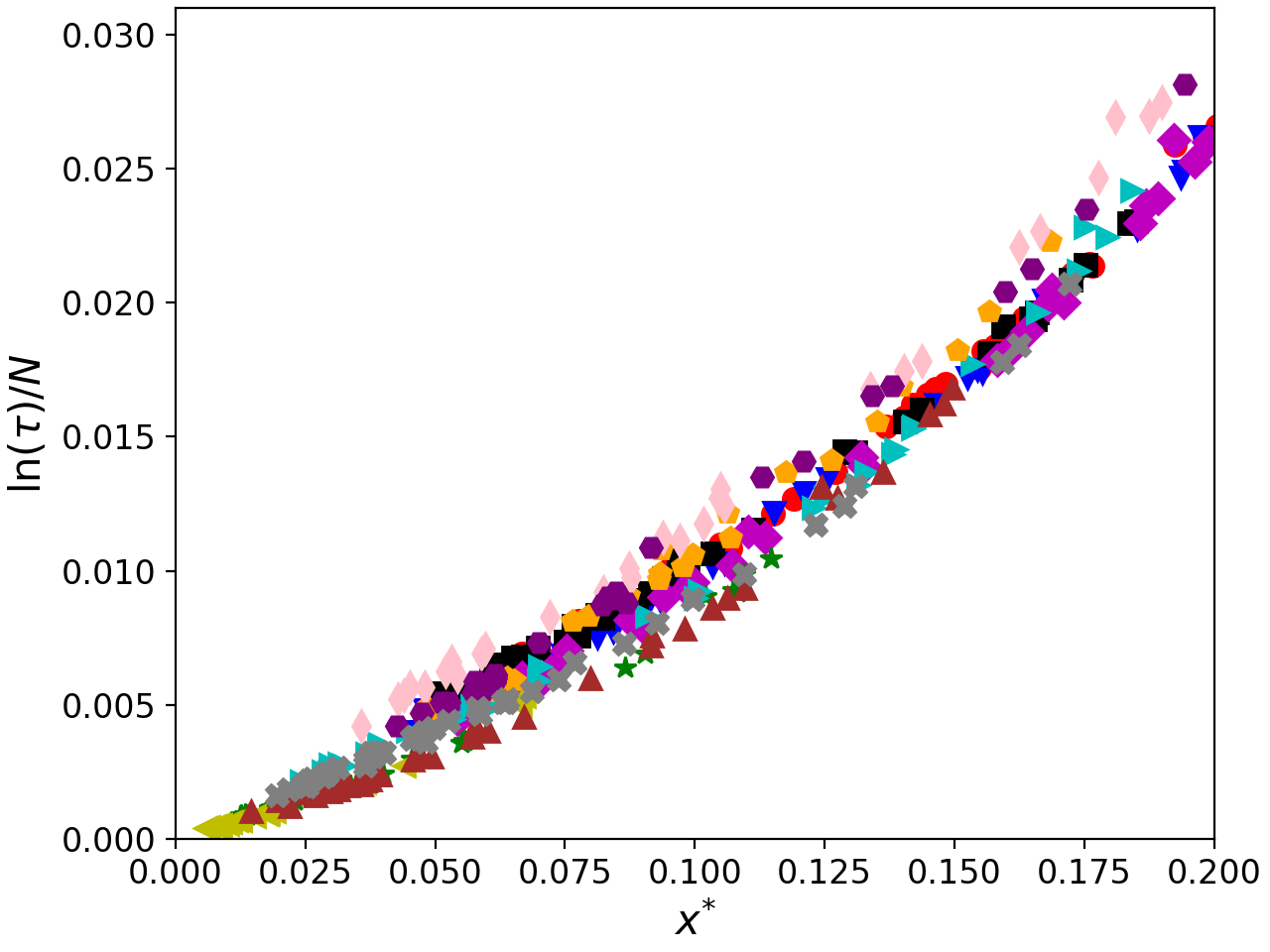}
\vspace{-4mm}    
\caption{The normalized logarithm of the MTE, $\ln(\tau)/N$, versus the disease prevalence $x^{*}$. Here,  twelve different symbols represent gamma, beta, log-normal and inverse-Gaussian degree distributions with network sizes ranging between $3\times10^3\leq N\leq 3\times 10^4 $ and  $1.09\leq R_0\leq1.36$. Each symbol represents a WE simulation with different degree heterogeneity and assortativity,  $\epsilon\leq3.0$ and $-0.6\leq\alpha\leq0.6$, which gives rise to different values of $x^{*}$. The latter is computed for each simulation as the mean of the obtained QSD, see also Fig.~\ref{fig3}.} 
    \label{fig5}
\end{figure}

The fact that $\epsilon$ and $\alpha$ are interchangeable, and that the MTE is largely determined by the disease prevalence is not limited to bimodal networks, and extends to generic heterogeneous networks, as presented in Fig.~\ref{fig5}. Here, we plot the logarithm of the MTE divided by the network size for various networks, with different degree distributions, while varying \(N\), \(R_0\), \(\epsilon\), and \(\alpha\), such that the disease prevalence \(x^*\) remains fixed. The figure shows that, when $x^*$ is not too large, irrespective of the degree distribution, assortativity, reproductive number, or network size, the quantity \(\ln(\tau)/N\) remains approximately constant as long as \(x^*\) is fixed. Combined with the results in Fig.~\ref{fig1}, showing that the action barrier satisfies \(\Delta\mathcal{S} \sim \ln(\tau)/N\) with logarithmic corrections, we conclude that the action barrier is almost entirely determined by the disease prevalence.  As computing \(x^*\) is a much simpler task than estimating the MTE via simulations, this result serves as a practical and efficient tool for estimating the MTE.

\begin{figure}[t]
    \includegraphics[width=0.91\linewidth]{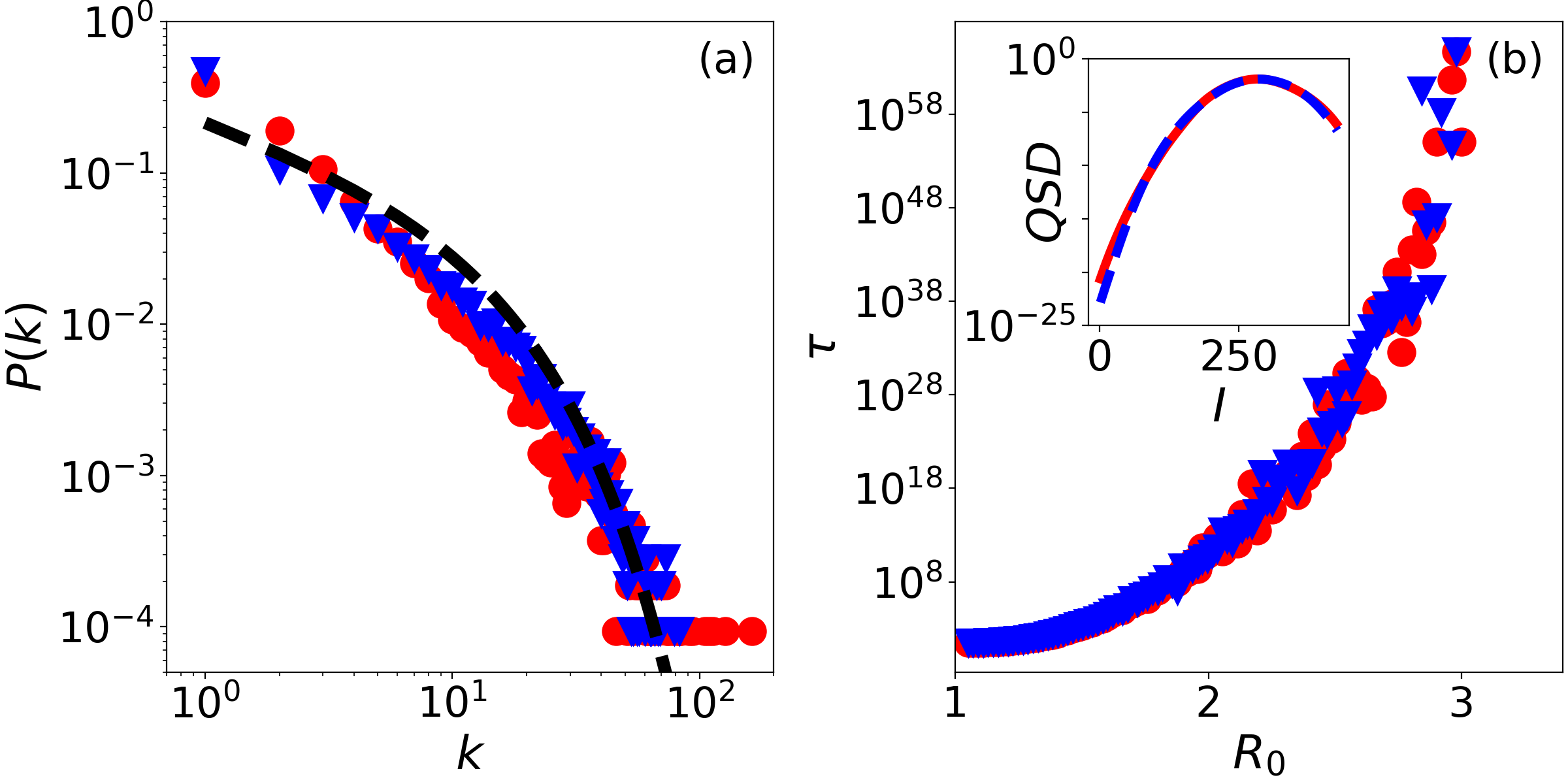}
\vspace{-5mm}    
\caption{(a) Degree distributions, \( P(k) \), for the PGP network (circles), see text, and for a realization of a gamma network (triangles), while  the dashed line denotes the theoretical gamma distribution, with \( N = 10680 \), \( \langle k \rangle = 4.55 \), and \( \epsilon = 1.77 \). (b) The MTE verus the basic reproductive number $R_0$. Here the circles and triangles are WE  simulation results for the same networks as in panel (a) with $\alpha=0.238$, and with WE parameters \( \tau_{WE} = 0.5 \), \( m = 500 \), and \( M = 70 \). The inset of panel (b) displays the QSD for the PGP network (solid line) compared to that of the gamma network (dashed line), for the same networks as in the main figure, with \( R_0 = 2.9 \).
} 
    \label{fig6}
\end{figure}

So far, we have studied the dependence of the disease lifetime on the degree heterogeneity and degree assortativity of \textit{synthetic} heterogeneous networks. At this point, we wanted to check whether realistic networks can also be studied using this formalism, and whether it is possible to mimic such networks by using synthetic networks with a presecribed degree distribution and degree-degree correlations. In order to do so, we took the so-called \textit{Pretty Good Privacy} (PGP) encryption network~\cite{Chakrabarti_2008_Epidemic_thresholds_in_real_networks}, which is an example of a real-world, or empirical network.  This network emerges when secure information is shared between members using the PGP encryption algorithm. Here, a pair of keys is made, a public key for encrypting messages and a private key for decrypting them. Individuals generate their own key pairs, publish the public key, and keep the private key secret. To communicate securely, the sender encrypts the message using the recipient's public key, while the recipient uses their private key to decrypt it. Thus, a "web of trust" is built, in which individuals validate each other’s encryption keys~\cite{Chakrabarti_2008_Epidemic_thresholds_in_real_networks}.  In Fig.~\ref{fig6}(a) we show the degree distribution $P(k)$ of the PGP network, compared with a gamma network with  parameters such that the mean and standard deviation of $P(k)$ match that of the PGP network. Furthermore, by tuning the assortative mixing of the gamma network such that it coincides with that of the PGP network and applying the SIS dynamics, we find that the MTE of both networks agree very well. This agreement demonstrates that the effect of degree heterogeneity and assortative mixing on rare events in population networks can be studied using synthetic networks, which can be tuned to fully mimic the topology of real-world networks.

\section{Discussion \label{sec:discussion}}
The study of disease spread in assortative networks has primarily centered on the dynamics of the endemic state and its associated threshold. Nevertheless, rare events such as disease extinction have received considerably less attention, due to the analytical complexity and  significant computational time. Here, we used numerical and analytical methods to study disease extinction in heterogeneous assortative population networks. Our numerical results were mainly obtained using the weighted ensemble (WE) method, while the analytical methods relied on  a semiclassical approximation to the  master equation.

Despite the high dimensionality of the problem, we managed to derive a simple analytical result for the mean time to extinction (MTE) of the disease in the regime of weak heterogeneity, characterized by a small coefficient of variation of the network's degree distribution. While our results coincide with existing results for random networks (with zero assortativity), our analytical findings allow to rigorously quantify the effect of assortative (or disassortative) mixing on the MTE. Importantly, we have demonstrated that increasing assortativity amplifies the frequency of rare events, leading to a shorter disease lifetime, similarly as the effect of increasing heterogeneity. All our analytical results were tested against extensive numerical WE simulations over a broad range of parameters. In this realm, the WE method proved to be a highly efficient tool for numerically estimating the MTE and determining the quasistationary distribution in parameter regimes, which were previously inaccessible using standard kinetic Monte-Carlo techniques.

One of the main findings of this work is the qualitative resemblance between the role of assortativity and degree heterogeneity, where these parameters can be balanced in such a way to maintain a fixed MTE. Remarkably, for low reproductive numbers, along these curves where the MTE is fixed we observed an almost constant disease prevalence in the quasi-stationary state, indicating that in such networks the MTE is primarily governed by the disease prevalence. Naturally, while deviations from this result occur when the disease prevalence becomes significant, this result is valid for a broad class of diseases with a relatively low $R_0$, such as seasonal influenze~\cite{biggerstaff2014estimates}. 

We also tested the WE method on a real-life network and showed that synthetic networks can be used to reliably predict the MTE, as long as the network topology matches that of the empirical network. While beyond the scope of this work, it would be interesting to also study the interplay between assortativity and the distribution's higher centralized moments such as skewness, and how their combined influence affects the MTE~\cite{Hindes_2016_jason-ira-paths,korngut_2024_we}. 

\section{Acknowledgments}
\vspace{-3mm}
\noindent
EK and MA acknowledge support from ISF grant 531/20.

\bibliography{bibfile}
\end{document}